\documentclass[manuscript]{aastex}

\newcommand{\CeighteenO}{C$^{18}$O}
\newcommand{\HthirteenCOp}{H$^{13}$CO$^{+}$} 
\newcommand{\NtwoHp}{N$_{2}$H$^{+}$}
\newcommand{\J}[2]{($J$=#1--#2)}
\newcommand{\Jno}[2]{$J$=#1--#2}
\newcommand{\HII}{H {\sc ii}}
\newcommand{\Msun}{M$_{\odot}$}
\newcommand{\pcc}{cm$^{-3}$}
\newcommand{\Mcore}{$M_{\rm core}$}
\newcommand{\Mvir}{$M_{\rm vir}$}
\newcommand{\Rcore}{$R_{\rm core}$}
\newcommand{\dvcore}{$dv_{\rm core}$}
\newcommand{\ncore}{$\bar{n}$}
\newcommand{\xe}[1]{$\times 10^{#1}$}
\newcommand{\sime}[1]{$\sim 10^{#1}$}
\newcommand{\kms}{km s$^{-1}$}
\newcommand{\Vlsr}{$\upsilon_{\rm{LSR}}$}
\newcommand{\degree}{$^{\circ}$}
\newcommand{\equatorial}[6]{#1$^{\rm h}$#2$^{\rm m}$#3$^{\rm s}$, #4\degree#5$'$#6$''$}

\newcommand{{\TAstar}}{$T_{\rm{A}}^{*}$}
\newcommand{\Resolution}[1]{$\Delta \theta_{\rm #1}$}

\shorttitle{C$^{18}$O CMF in the S 140 region}
\shortauthors{Ikeda \& Kitamura}

\begin{document}

\title{Similarity between the {\CeighteenO} \J{1}{0} core mass function and the IMF in the S 140 region}

\author{Norio Ikeda}
\affil{Institute of Space and Astronautical Science/Japan Aerospace Exploration Agency, 
3-1-1 Yoshinodai, Chuo, Sagamihara, Kanagawa 229-8510, Japan}
\email{nikeda@ir.isas.jaxa.jp}

\and

\author{Yoshimi Kitamura}
\affil{Institute of Space and Astronautical Science/Japan Aerospace Exploration Agency, 
3-1-1 Yoshinodai, Chuo, Sagamihara, Kanagawa 229-8510, Japan}

\begin{abstract}
We present the results of {\CeighteenO}\J{1}{0} mapping observations
of a $20'\times18'$ area in the Lynds 1204 molecular cloud associated with
the Sharpless 2-140 (S140) {\HII} region.
The {\CeighteenO} cube ($\alpha$-$\delta$-$v_{\rm LSR}$) data shows that
there are three clumps with sizes of $\sim$ 1 pc in the region.
Two of them have peculiar red shifted velocity components at their edges,
which can be interpreted as the results of
the interaction between the cloud and the Cepheus Bubble.
From the {\CeighteenO} cube data,
the {\tt clumpfind} identified 123 {\CeighteenO} cores,
which have mean radius, velocity width in FWHM, and
LTE mass of 0.36$\pm$0.07 pc, 0.37$\pm$0.09 {\kms}, and
41$\pm$29 {\Msun}, respectively.
All the cores in S140 are most likely to be gravitationally bound
by considering the uncertainty in the {\CeighteenO} abundance.
We derived a {\CeighteenO} core mass function (CMF),
which shows a power-law-like behavior above a turnover at 30 {\Msun}.
The best-fit power-law index of $-2.1\pm0.2$ is quite
consistent with those of the IMF and
the {\CeighteenO} CMF in the OMC-1 region by \citet{ike09b}.
\citet{kra98} estimated the power-law index of $-$1.65 in S140
from the {\CeighteenO}\J{2}{1} data,
which is inconsistent with this study.
However,
the {\CeighteenO}\J{2}{1} data are spatially limited to the central part of the cloud
and are likely to be biased toward high-mass cores,
leading to the flatter CMF.
Consequently, this study and our previous one strongly support that
the power-law form of the IMF has been already determined
at the density of 
$\lesssim$ $10^{3\mbox{\scriptsize --}4}$ {\pcc},
traced by the {\CeighteenO}\J{1}{0} line.
\end{abstract}




\section{Introduction}
\label{introduction}
It is believed that new stars are born in dense molecular cloud cores.
Therefore, the physical properties of the dense cores are thought to be
closely related to the masses of new stars to form within them.
Particularly, the dense core mass function (DCMF) has been considered
as a key to understanding the stellar initial mass function (IMF).
Many authors have investigated the DCMFs
by using dense ($\geq 10^{4\mbox{\scriptsize --}5}$ {\pcc}) gas tracers such as
(sub)millimeter dust continuum emission, infrared visual extinction, and
molecular line emissions having high critical densities, for example,
the {\HthirteenCOp} \J{1}{0} and {\NtwoHp} \J{1}{0} lines
\citep[e.g.,][]{mot98,rei06,nut07,rat09,ike07,wal07,eno08}.
In nearby ($\leq$ 500 pc) star forming regions 
such as Orion, Ophiuchus, the Pipe Nebula, Perseus, and Serpens,
they found that the DCMFs seem to have power-law-like behaviors in high-mass parts
as $dN/dM \propto M^{-\gamma}$,
whose $\gamma$ values are very similar to that of the IMF.
Based on these observational facts,
they propose a hypothesis that
the power-law form of the IMF has been already determined at the formation stage of the dense cores.

\citet{ike09b} have recently discovered a similarity 
between the power-law forms of
the IMF and the core mass function (CMF)
even in the tenuous 
($10^{3\mbox{\scriptsize --}4}$ {\pcc})
cloud structures.
They carried out mapping observations by using the {\CeighteenO} \J{1}{0} line,
whose critical density is 2\xe{3} {\pcc} \citep{yos10},
toward the OMC-1 region in the Orion A cloud.
They found
the $\gamma$ value of the {\CeighteenO} CMF of 2.3 -- 2.4.
The $\gamma$ value is not only similar to those of 
the H$^{13}$CO$^{+}$ DCMF  \citep[$2.2\pm0.1$;][]{ike07}
and the 850 {\micron} dust continuum DCMF  \citep[2.2$\pm$0.2;][]{nut07} within the uncertainties,
but also quite consistent with that of the IMF of the Orion Nebula Cluster \citep[2.2;][]{mue02},
which is associated with the OMC-1 region.
The agreement between the C$^{18}$O CMF and IMF $\gamma$ values
suggests that,
at least in the OMC-1 region,
the power-law form of the IMF with $\gamma \geq 2$
has been maintained from the formation time of
the tenuous structure 
with density of
\sime{3\mbox{\scriptsize --}4} {\pcc}.
In addition to the Orion cloud, a representative of massive star formation,
the similarity between the CMF and the IMF was reported
in low-mass star forming regions by \citet{tac02}.
The {\CeighteenO} cores found in Taurus, Chamaeleon, Lupus, and other low-mass star forming regions
show the CMF with $\gamma$ = 2.6 above 56 {\Msun}.

\citet{kra98}, however, found a significantly smaller $\gamma$ value of 1.7 for the S140 and M17SW regions
by using the {\CeighteenO} \J{2}{1} line
with a high critical density comparable to that of {\HthirteenCOp} \J{1}{0}.
\citet{won08} also found $\gamma$ = 1.7 in the {\CeighteenO} \J{1}{0} CMF of RCW 106.
In this study,
we focus on S140 for the following reasons.
First, the distance to S140 of 910 pc \citep{cra74} is the nearest one of them;
the distances to M17SW and RCW 106 are 2.2 kpc \citep{chi80} and 3.6 kpc \citep{loc79}, respectively,
and the two clouds are located in the Sagittarius arm.
Since we discuss the power-law nature of the CMF
on the basis of a comparison with that in OMC-1,
it seems quite natural to first select the S140 region,
located in the Local (Orion) Spur as the next step in our CMF study.
Second,
since the previous {\CeighteenO} \J{2}{1} observations of S140 \citep{joh92} 
covered only the brightest $4'\times4'$ region around IRAS 22176+6303,
we cannot exclude the possibility that
their core identification was biased to high-mass cores, leading to the flatter CMF.
%
Third,
the higher transition of \Jno{2}{1} with
the transition energy of 15.8 K might prefer to pick up higher-mass protostellar cores,
compared to the lowest transition of \Jno{1}{0}.

The aim of this paper is to re-estimate the power-law index of the CMF in the S140 region,
also identified as Lynds 1204,
and to examine whether the similarity between the CMF and the IMF holds in the region or not.
Our observations were done to cover the cloud as widely as possible,
in order to avoid the possible spatial bias.
We employed the {\CeighteenO} \J{1}{0} line emission,
which is the same tracer as \citet{ike09b} did, and
is suitable for deriving the tenuous CMF \citep[e.g.,][]{tac02}.
Furthermore,
we used the {\tt clumpfind} algorithm \citep{wil94}, the same core identification method as that \citet{ike09b} employed,
to directly compare with their results.
Note that the spatial resolution is high enough to evaluate the CMF 
at the distance to the S140 region of 910 pc, as shown in \S \ref{mf_text}.

In \S \ref{observation}, we describe the details of our observations.
In \S \ref{results}, we show the overall spatial and velocity structures of
the tenuous gas traced by the {\CeighteenO} \J{1}{0} emission in the S140 cloud,
and identify three clumps with sizes of $\sim$ 1 pc.
In \S \ref{catalog} 
we describe the identification of the {\CeighteenO} cores in S140, and
discuss their physical properties by comparing with those in the OMC-1 region.
We show the {\CeighteenO} \J{1}{0} CMF in \S \ref{mf_text}.
In this section,
we compare our CMF with the previous \Jno{2}{1} CMF,
and discuss the similarity between the {\CeighteenO} (1--0) CMF and the IMF.
In \S \ref{summary} we summarize our results.

\section{Observations}
\label{observation}
The {\CeighteenO} \J{1}{0} mapping observations were carried out 
in the period from January to February 2010
by using the Nobeyama 45 m radio telescope.
Our map covered a $20'\times18'$ area,
whose center was selected to be \equatorial{22}{20}{18}{63}{22}{12} (J2000).
Using the On-The-Fly technique \citep{saw08},
we swept the area by raster scan
with a scan speed of the telescope of 35 arcsec s$^{-1}$.
To reduce scanning effects, 
we scanned in both the RA and Dec directions.
At the frequency of the {\CeighteenO} \J{1}{0} emission
\citep[109.782182 GHz;][]{ung97},
the half power beam width, \Resolution{HPBW},
and main beam efficiency, $\eta$, of the telescope were 
14$''$ and 0.4, respectively.
By the standard chopper-wheel method,
the receiver intensity was converted into the antenna temperature {\TAstar},
corrected for the atmospheric attenuation.
We used an off position of \equatorial{22}{18}{34}{63}{2}{12},
where we could not the {\CeighteenO} emission above 0.09 K in {\TAstar}.
At the front end, we used the 25-BEam Array Receiver System (BEARS) in double-sideband (DSB) mode, 
which has 5$\times$5 beams separated by 41$''$.1
in the plane of the sky \citep{sun00, yam00} .
The 25 beams have beam-to-beam variations of about 10\%
in both beam efficiency and sideband ratio.
To correct for the beam-to-beam gain variations,
we calibrated the intensity scale of each beam
by observing the W3 region
with a 100 GHz SIS receiver (S100) with a single-sideband filter.
At the back end, we used
25 sets of 1024 channel autocorrelators (ACs),
which have a velocity resolution
of 0.104 km s$^{-1}$ at 110 GHz \citep{sor00}.
Since the data dumping time of the ACs was 0.1 s,
the spatial data sampling interval on the sky plane was 3.5$''$.
The spatial interval corresponds to 0.3\Resolution{HPBW},
and is small enough to satisfy the Nyquist theorem.
The telescope pointing was checked every 1.5 hours
by observing the SiO ($v$=1, $J$=1--0; 43.122 GHz) maser source T--Cep.
Since the pointing uncertainty of the telescope remains as small as a few arcseconds
below a wind speed of 5 m s$^{-1}$,
we rejected the data taken under the condition that
the wind speed averaged over one minute exceeds 6 m s$^{-1}$.
Consequently, the pointing accuracy is better than 3$''$
for the data to be analyzed.

To construct an $\alpha$-$\delta$-$v_{\rm LSR}$ data cube
with spatial and velocity resolutions of 22$'' $ and 0.1 {\kms} in full width half maximum (FWHM),
we used a Gaussian function as a gridding convolution function (GCF)
to integrate the spectra which were taken
with the very high spatial sampling rate of 3.5$''$.
We adopted 17$''$.0 as the size of the GCF 
in FWHM.
The resultant effective spatial resolution of the cube,
\Resolution{eff} becomes 22$''$.0,
corresponding to 0.1 pc at the distance to the S140 region.
Since the DSB system noise temperature ranged from 285 to 357 K
with a mean value of 321 K during the observations,
we have an rms noise level of the data cube of 0.12 K in {\TAstar}.

\section{{\CeighteenO} \J{1}{0} map of the S140 region}
\label{results}
In this section we describe the overall spatial and 
velocity structures of the {\CeighteenO} \J{1}{0} emission in the S 140 region.
In addition, we identify three distinct clumps with sizes of $\sim$ 1 pc,
which is likely to be the natal objects for clusters.

\subsection{Spatial and Velocity Structures of the {\CeighteenO} \J{1}{0} emission}
\label{map}
\begin{figure}
\epsscale{1}
\plotone{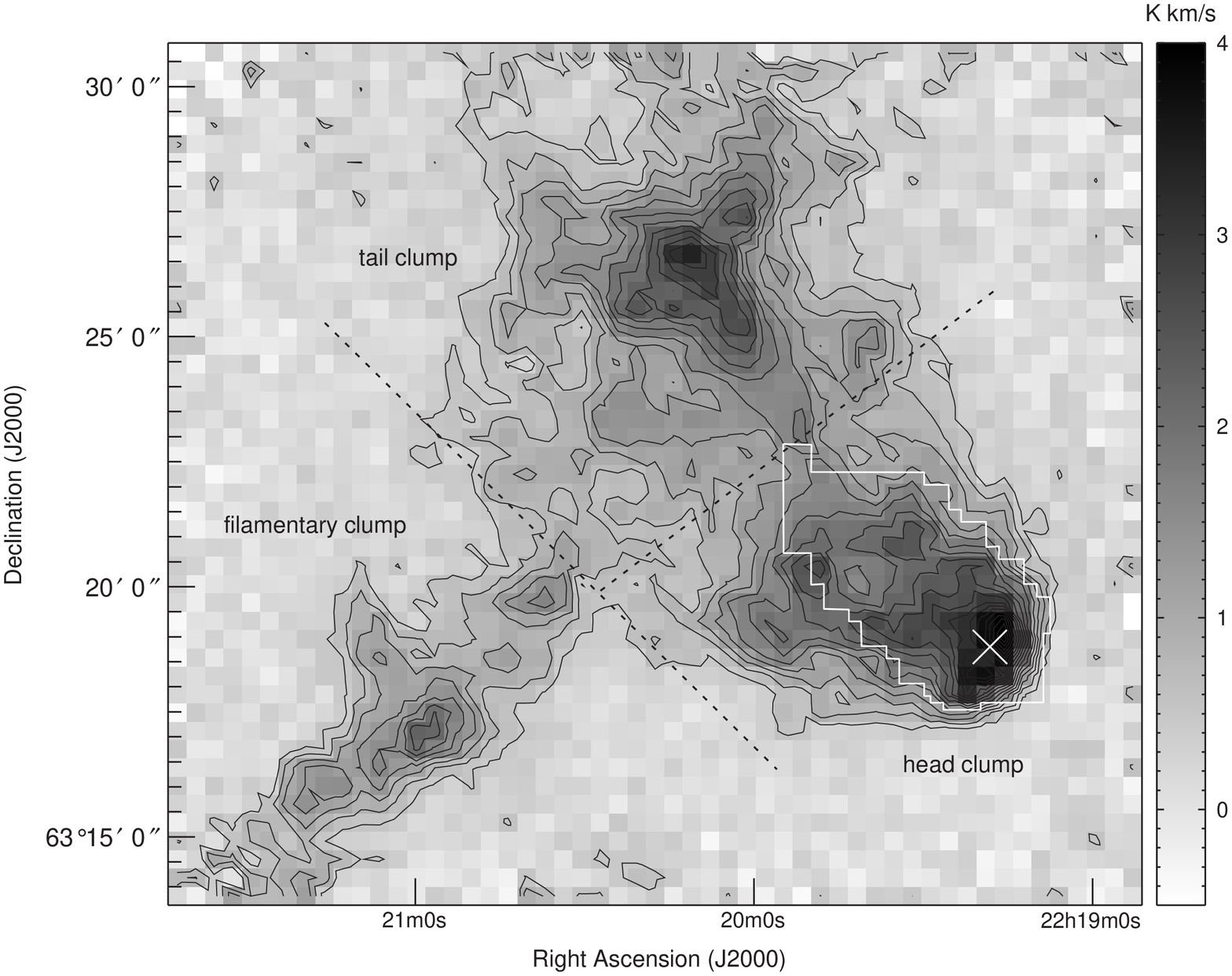}
\caption{Total integrated intensity map of the {\CeighteenO} \J{1}{0} emission
( \Vlsr = $-9.9$ - $-4.8$ {\kms})
in the S 140 region.
The contour intervals are 0.25 K {\kms} (corresponding to 2$\sigma$) 
starting at 0.5 K {\kms}.
The gray scale bar is shown at the right-hand side of the panel.
The white cross indicates the position of IRAS 22176+6303.
The white polygon shows the mapping region in {\CeighteenO} \J{2}{1} by \citet{joh92}.
The dashed lines roughly indicate the boundaries of
the head, tail, and filamentary clumps (see \S \ref{map}).
}
\label{totalMap}
\end{figure}

Figure \ref{totalMap} shows the total integrated intensity map of
the {\CeighteenO} \J{1}{0} emission of the S140 region.
The total mass of the cloud is estimated to be 6600 {\Msun} (see \S \ref{coreidentification} for details).
One can see three distinct peaks, referred to as head, tail and filamentary clumps in this paper.
The main body of the cloud is known as a cometary one, and our map clearly shows the cometary shape;
the {\CeighteenO} emission is the brightest toward the head clump,
which faces the Sh-2 140 H {\sc ii} region and
is associated with IRAS 22176+6303, indicated by the cross mark in Figure \ref{totalMap}.
The head clump is also known as an active cluster-forming region
where numerous young stellar objects have been identified in infrared wavelength \citep{meg04}.
On the other hand, 
the tail clump seems streaming toward the north-east.
In contrast to the head clump,
no active star formation is known within the tail clump,
although no sensitive infrared observations have been published yet.
Since our map covers a wider region by a factor of three than
those of the previous molecular line mapping observations \citep{rid03,hig09},
we found another filamentary clump located on the south-east of the cometary cloud.
Note that
the elongation of the filamentary clump seems perpendicular to the head-tail direction of the cometary cloud.
Although the filamentary clump can be seen
in a previous large-scale survey with the spatial resolution of several arcminutes \citep{yon97},
our map is the finest one resolving 0.1 pc-scale cores.

\begin{figure}
\epsscale{1}
\plotone{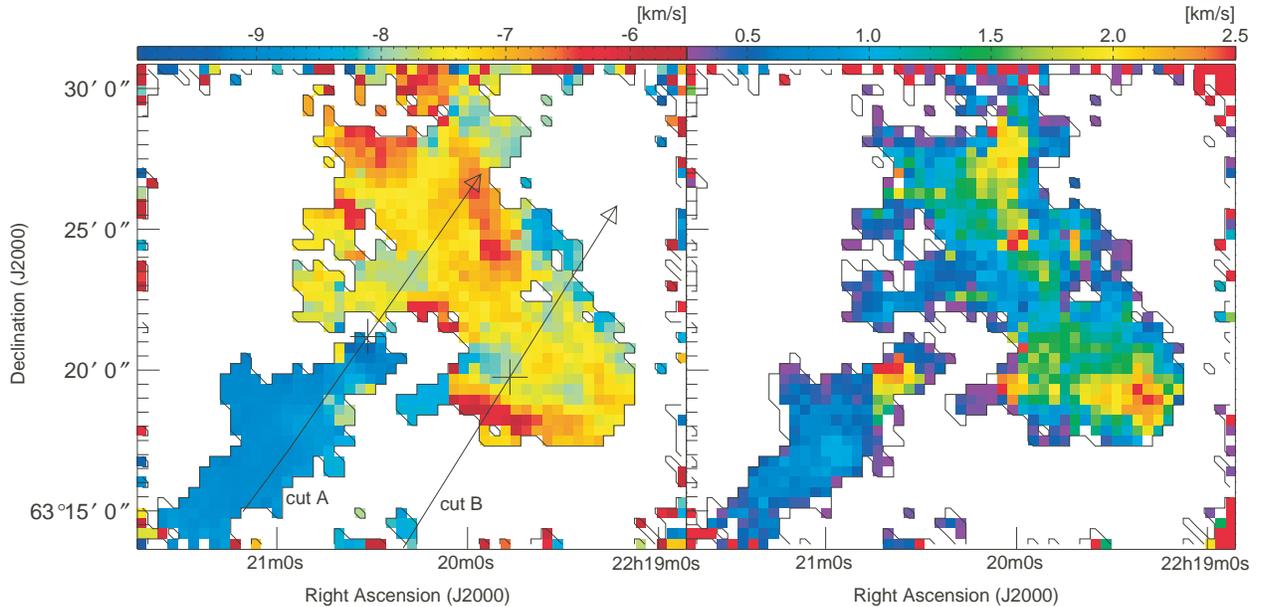}
\caption{
({\it Left}) Map of the intensity-weighted mean LSR velocity, i.e., the centroid velocity map
for the {\CeighteenO} \J{1}{0} emission in the S 140 region.
({\it Right}) Map of the velocity width in FWHM.
The color bar is shown at the top of each panel.
In the left panel,
the arrows indicate the cutting lines of the position-velocity diagrams in Figure \ref{pv}.
The cross marks on the arrows show the origins of the position axis for the position-velocity diagrams.
The centroid velocity and the velocity width are calculated within the areas above the 3 $\sigma $ noise level:
the thin solid polygons delineate the 3 $\sigma $ noise level
in Figure \ref{totalMap} (see also \citet{yos10}).
}
\label{velocityField}
\end{figure}

\begin{figure}
\epsscale{1.0}
\plotone{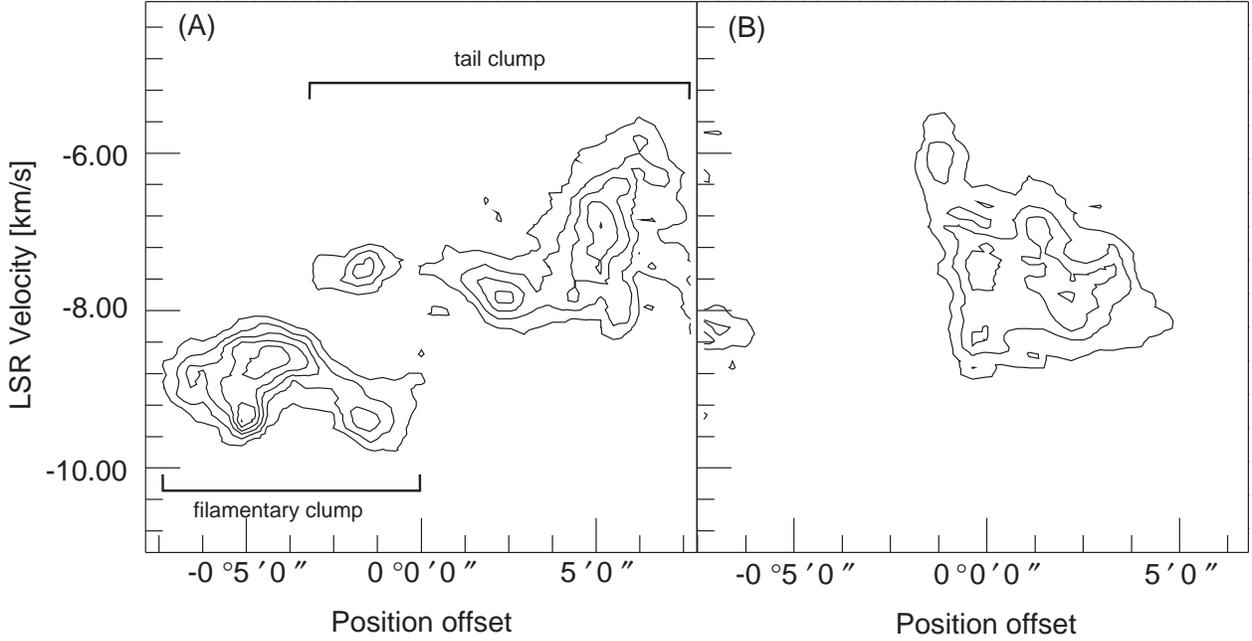}
\caption{Position-velocity diagrams of the {\CeighteenO} \J{1}{0} emission
along the cutting lines A ({\it left}) and B ({\it right})
indicated in Figure \ref{velocityField}.
The origins of the position axes are indicated by the cross marks in Figure \ref{velocityField}.
The contour intervals are 0.25 K (corresponding to 2$\sigma$) 
starting at 0.38 K (3 $\sigma$).
}
\label{pv}
\end{figure}

As shown in Figure \ref{velocityField}, 
we calculate the intensity-weighted mean LSR velocity, i.e., the centroid velocity, and
velocity width in FHWM of the {\CeighteenO} emission at each spatial grid point
in Figure \ref{totalMap}, following \citet{yos10}.
The centroid velocity map illustrates that
the centroid velocities in the cometary cloud range from $-8.5$ to $-6.0$ {\kms},
while the filamentary clump has a more blue-shifted velocity range from $-10.0$ to $-8.0$ {\kms}.
The mean velocity width all over the cometary cloud is 1.0 {\kms},
and the largest velocity width of 2.5 {\kms} is found toward the IRAS source,
which is probably related to the star formation activity in the head clump.
On the other hand,
the velocity width of the filamentary clump is as small as 0.6 {\kms}.
However,
the velocity width map apparently shows
a somewhat large velocity width of $\sim$ 2.5 {\kms}
at $\alpha, \delta$ = \equatorial{22}{20}{43}{63}{20}{0} in the filamentary clump. 
As indicated in the position-velocity diagram of Figure \ref{pv}A,
this is because
a component belonging to the cometary cloud with a mean LSR velocity of $\sim$ $-7.5$ {\kms} overlaps
with the filamentary clump with a mean LSR velocity of $\sim$ $-9.0$ {\kms} along the line of sight.
Note that the velocity width of each components is as small as $\sim$ 0.5 {\kms} at this point.


The centroid velocity map indicates that
there are red-shifted components of {\Vlsr} = $-7 \: \sim \: -6$ {\kms} on the periphery of the cometary cloud.
Among the red-shifted components, the most prominent one can be found in the southern edge of the head clump.
The red-shifted component of $-6$ {\kms}, for example, is more easily recognized
in the position-velocity diagram of Figure \ref{pv}B.
Since the S140 region faces the Cepheus Bubble and the cloud exhibits the cometary shape,
the region is likely to interact with the expanding bubble driven by
stellar winds and supernova explosions of members of Cep OB2 \citep{abr00}.
Therefore,
the red-shifted components around the cometary cloud can be interpreted as
the outer envelope of the cloud
that has been pushed toward the far side along the line of sight by the expanding bubble.
However,
the velocity widths of the red-shifted components of $\sim 0.5$ {\kms}
never exceed that of the main body of the cometary cloud.
Since the expanding velocity of the bubble of 10 {\kms} \citep{abr00}
is an order of magnitude larger than the thermal motion of the cloud,
it is expected that the interaction is associated with shocks,
leading to large velocity width.
The small velocity widths of the red-shifted components might imply that
the turbulent motions
excited by the interaction in the red-shifted components have been already dissipated
with a crossing time as short as 7\xe{4} yr
(see also a discussion in \S \ref{coreProperties}).


\subsection{Identification of Clumps for Cluster Formation}
\label{clumpProperties}

\citet{lad03} showed that
the clumps with sizes of $\sim$ 1 pc and masses of the orders of $10^2$ -- $10^3$ {\Msun} are
the natal objects for the stellar clusters.
The three clumps in S140, the head, tail and filamentary clumps can be considered as
the cluster-forming clumps, though active cluster formation has been found only in the head clump.
This is because their sizes are about 1 pc and the masses are of the order of $10^{3} $ {\Msun}.
Actually, the masses of the head, tail, and filamentary clumps are
estimated to be 2200, 2600, and 1800 {\Msun}, respectively,
from the total  integrated intensities with {\Vlsr} = $-$9.9 -- $-4.8$ {\kms}
of the areas above the 2 $\sigma$ noise level in Figure \ref{totalMap}
(the details of the mass estimation is shown in \S \ref{coreidentification}).
Previously, \citet{rid03} and \citet{hig09} mapped the head clump by using the {\CeighteenO} \J{1}{0} line,
and the clump mass was estimated to be 1900 {\Msun} by \citet{hig09},
which is roughly consistent with ours.


Our interpretation that the three clumps in S140 are the site of cluster formation
is also supported by the following virial analysis.
Assuming the clumps are spherical, we derived the virial masses of 1100, 1300, 330 {\Msun}
with the uncertainty of a factor of 3
for the head, tail, and filamentary clumps, respectively.
Since the clump masses are twice or more larger than the virial masses,
it is most likely that the clumps are gravitationally bound 
and have the potential for producing clusters.
In addition, the whole cloud of S140 is also in gravitationally bound state.
By using
the mean velocity width of 1.95 {\kms} in FWHM all over the cloud and
the projected extent of the cloud of 3.7 pc of the cloud,
the virial mass of the cloud is estimated to be 1900 {\Msun},
which is by a factor of three smaller than the cloud mass of 6600 {\Msun}.

\section{{\CeighteenO} core catalog}
\label{catalog}

\subsection{Identification of the Cores and Derivation of Their Physical Properties}
\label{coreidentification}
Following the {\CeighteenO} CMF study in the OMC-1 region \citep{ike09b},
we applied the {\tt clumpfind} algorithm \citep{wil94}
to the {\CeighteenO} three-dimensional ($\alpha$-$\delta$-$v_{\rm LSR}$) cube data.
The algorithm can work well with reasonable parameters
to identify cores or clumps, though several authors pointed out some
shortcomings of the {\tt clumpfind}.
\citet{pin09} examined
the behavior of the algorithm by changing the threshold level from 3 to 20 $\sigma$,
a wider range than \citet{wil94} examined,
and found that the power-law index of the mass function sensitively depends on
the threshold for higher thresholds of $>$ 5 $\sigma$.
However,
\citet{ike09b} demonstrated
the weak dependence of core properties and CMF
on the threshold in the reasonable range from 2 to 5 $\sigma$ levels,
which had been also shown in \citet{pin09}.
Therefore, 
we adopted the threshold level for the algorithm of 0.24 K,
i.e., the 2 $\sigma$ noise level of the cube data,
which is recommended for identifying the core structure by \citet{wil94}
and falls in the robust and reasonable range derived by \citet{ike09b}.
We adopted the grid spacing of the cube data of 22$''$.0,
equal to \Resolution{eff},
i.e., full-beam sampling. 
This is because \citet{wil94} determined the optimal threshold of the 2 $\sigma$ level
for the full-beam sampling case.
We also used the additional criteria introduced in \citet{ike07}
to reject ambiguous or fake core candidates
whose size and velocity width
are smaller than the spatial and velocity resolutions, respectively.

We identified 123 cores and
estimated the beam-deconvolved radius {\Rcore},
velocity width in FWHM corrected for the spectrometer resolution {\dvcore},
LTE mass {\Mcore}, virial mass {\Mvir}, and mean density {\ncore} of the {\CeighteenO} cores.
Table \ref{propertyTable} shows the physical properties of the {\CeighteenO} cores in S140.
The definitions of these parameters are the same as those in \citet{ike09b}.
Here we briefly summarize the parameters specific to this study.
We adopted  \Resolution{eff} $= 22''.0$ as described in \S \ref{observation},
the antenna efficiency $\eta$ of 0.4,
and the fractional abundance of C$^{18}$O relative to H$_{2}$, $X_{\rm C^{18}O}$ of
$1.7\times10^{-7}$ \citep{fre82}.
The optical depth of the {\CeighteenO} line in the S140 region has
been estimated to be smaller than 1;
\citet{hig09} derived the upper limit of the optical depth of 0.5,
from the intensity ratio of the {\CeighteenO} \J{1}{0} to C$^{17}$O \J{1}{0} lines.
We assumed that the excitation temperature $T_{\rm ex}$ is uniform over the S140 region
and is equal to the rotational temperature of 24 K in the NH$_{3}$ (1, 1) and (2, 2) observations
by \citet{hig09}.
In the head clump,
the temperature of $>$ 20 K is reasonable
because the clump faces the Sh 2-140 H {\sc ii} region and
numerous young stellar objects have been found \citep{meg04},
as well as the OMC-1 cloud.
On the other hand, the other two clumps have not been well studied compared to the head clump.
In the filamentary clump, two IRAS point sources 22192+6302 and 22196+6302 are detected,
but the nature of the sources is unknown.
Toward the tail clump, no signature of star formation has been found.
Although our assumption of the uniform temperature cannot be validated for the two clumps,
we found that a low $T_{\rm ex}$ value of 10 K does not seriously affect
our discussion,
and therefore we adopted the assumption of the uniform $T_{\rm ex}$ in this study.

\subsection{Physical Properties of the {\CeighteenO} Cores}
\label{coreProperties}
We discuss the physical properties of the {\CeighteenO} cores 
on the basis of comparison with those of the {\CeighteenO} cores
in the OMC-1 region of the Orion A cloud \citep{ike09b}.
To fairly compare the S140 {\CeighteenO} cores with the OMC-1 {\CeighteenO} ones,
we smoothed the OMC-1 data by a Gaussian function with a FWHM size of 37$''$.3
so that the OMC-1 cloud is put at the distance of 910 pc to S140.
Furthermore, we randomly picked up 25 \% of the original spectra 
to make the cube data for OMC-1 so as to have the same signal to noise ratio as
that of the S140 data. 
Finally we re-identified 44 {\CeighteenO} cores from the smoothed cube data in OMC-1.

\begin{figure}
\epsscale{1.0}
\plotone{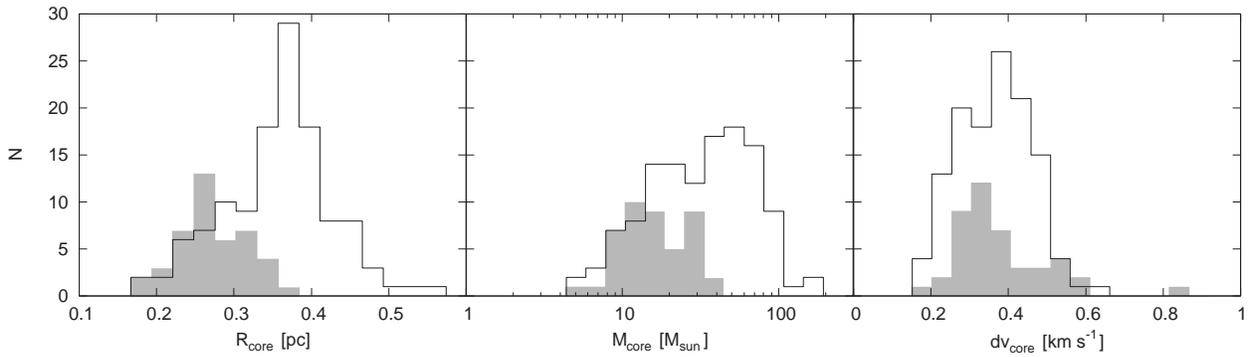}
\caption{
Histograms of {\Rcore} ({\it left}), {\Mcore} ({\it middle}),
and {\dvcore} ({\it right}) of the {\CeighteenO} cores.
The open and gray histograms mean the {\CeighteenO} cores in the S 140 region
and those in the smoothed OMC-1 data (see \S \ref{coreProperties}), respectively.
}
\label{histograms}
\end{figure}

The left panel of Figure \ref{histograms} shows that
the {\Rcore} distribution of the {\CeighteenO} cores in S140 has a single peak at 0.37 pc,
which is much larger than the peak radius of 0.26 pc in OMC-1:
the mean value of 0.36$\pm$0.07 pc for {\Rcore} in S140 is larger than that in OMC-1 of 0.27$\pm$0.05 pc.
The Kolmogorov-Smirnov (K-S) test applied to the {\Rcore} histograms demonstrates that
the distributions are considerably different from each other
with a significance level of 1 \%.
The mass {\Mcore} of the {\CeighteenO} cores in the S140 region tends to be also larger than
that in OMC-1,
as shown in the middle panel of Figure \ref{histograms};
the mean value in S140 of 41$\pm$29 {\Msun} is
twice larger than that in OMC-1 of 18$\pm$9 {\Msun}.
In addition,
the K-S test shows that the two distributions are considerably different from each other.
Although only one broad peak at $\sim$ 20 {\Msun} seems to exist for OMC-1,
we can see two broad peaks at $\sim$ 20 {\Msun} and $\sim$ 50 {\Msun} for S140.

The above comparisons suggest that
the large ({\Rcore} $>$ 0.4 pc) and massive ({\Mcore} $>$ 50 {\Msun}) cores
exist only in the S140 region.
Furthermore, the mass fraction of the cores in S140 is very large:
the total mass of the {\CeighteenO} cores is 5000 $M_{\odot}$,
75\% of the total mass traced by the {\CeighteenO} emission (see \S \ref{clumpProperties}). 
This fraction is twice larger than
that of the {\CeighteenO} cores in OMC-1 \citep{ike09b},
but is roughly consistent with the fraction of 60 \% for the {\HthirteenCOp} cores \citep{ike07,ike09}.
The large massive cores and the high mass fraction of the cores in the S140 region
are likely to be caused by high column densities in the region.
Actually, the {\CeighteenO} \J{1}{0} peak intensity of 2.4 K in {\TAstar} in S140 is
significantly larger than that in the OMC-1 region of 1.6 K.
It is possible to interpret that
the accumulation of interstellar matter occurred
owing to the interaction between the S140 cloud and the expanding Cepheus Bubble,
as suggested by the cometary shape and velocity features of the cloud as described in \S \ref{map}.

The {\ncore} range of 1.0 to 9.8\xe{3} {\pcc} for the {\CeighteenO} cores in S140
is similar to that in OMC-1 of 2.3 to 9.3\xe{3} {\pcc},
and is consistent with
the critical density of the {\CeighteenO} \J{1}{0} line of $\sim$ 2\xe{3} {\pcc} \citep{yos10}.
In addition,
\citet{rat09} showed that
the mean volume density of their {\CeighteenO} cores in the Pipe nebula is 7.3\xe{3} {\pcc},
which falls in our {\ncore} ranges of (1.0--9.8)\xe{3} {\pcc}.
In spite of the large range of {\ncore}, the traced
the traced densities of $\lesssim$ 10$^{4}$ {\pcc}
is significantly smaller than those of
the dense gas tracers used by the previous CMF studies mentioned in \S \ref{introduction}.

In contrast to {\Rcore} and {\Mcore},
the distribution of the velocity width {\dvcore} in S 140 is quite similar to that in OMC-1,
as shown in the right panel of Figure \ref{histograms}.
The K-S test shows that there is no considerable difference with a significance level of 1 \%.
The mean value of {\dvcore} in S 140 is 0.37$\pm$0.09 {\kms}
and is almost the same as that in OMC-1 of 0.38$\pm$0.12 {\kms}.
It is likely that the turbulent motions excited by the passing of the expanding bubble have been already damped,
though the bubble could significantly increase the velocity width of the {\CeighteenO} cores.
Considering the total kinetic energy of the bubble is 2.7\xe{50} erg in H {\sc i} gas,
the distance to the bubble center is 910 pc,
and the radius of the bubble is 5 degree \citep{abr00},
the kinetic energy injected into the {\CeighteenO} cores with a mean radius of 0.36 pc
can be estimated to be 1.7\xe{46} erg,
leading to a large velocity width of 1.2 {\kms};
such the large value of {\dvcore} cannot be discerned in the right panel of Figure \ref{histograms}.
Here, we assume that
the fraction of the injected kinetic energy to be converted into turbulence is
0.01 to 0.05 \citep{mac04,ike07}.
Since the expansion velocity of the bubble is 10.2 {\kms} \citep{abr00},
the crossing time of the bubble over the cores is estimated to be as short as
7\xe{4} yr, much smaller than the age of 1.7 Myr for the bubble \citep{abr00}.
Therefore, one can expect the dissipation of the turbulent motions within the cores at the present time.

\begin{figure}
\epsscale{1.0}
\plottwo{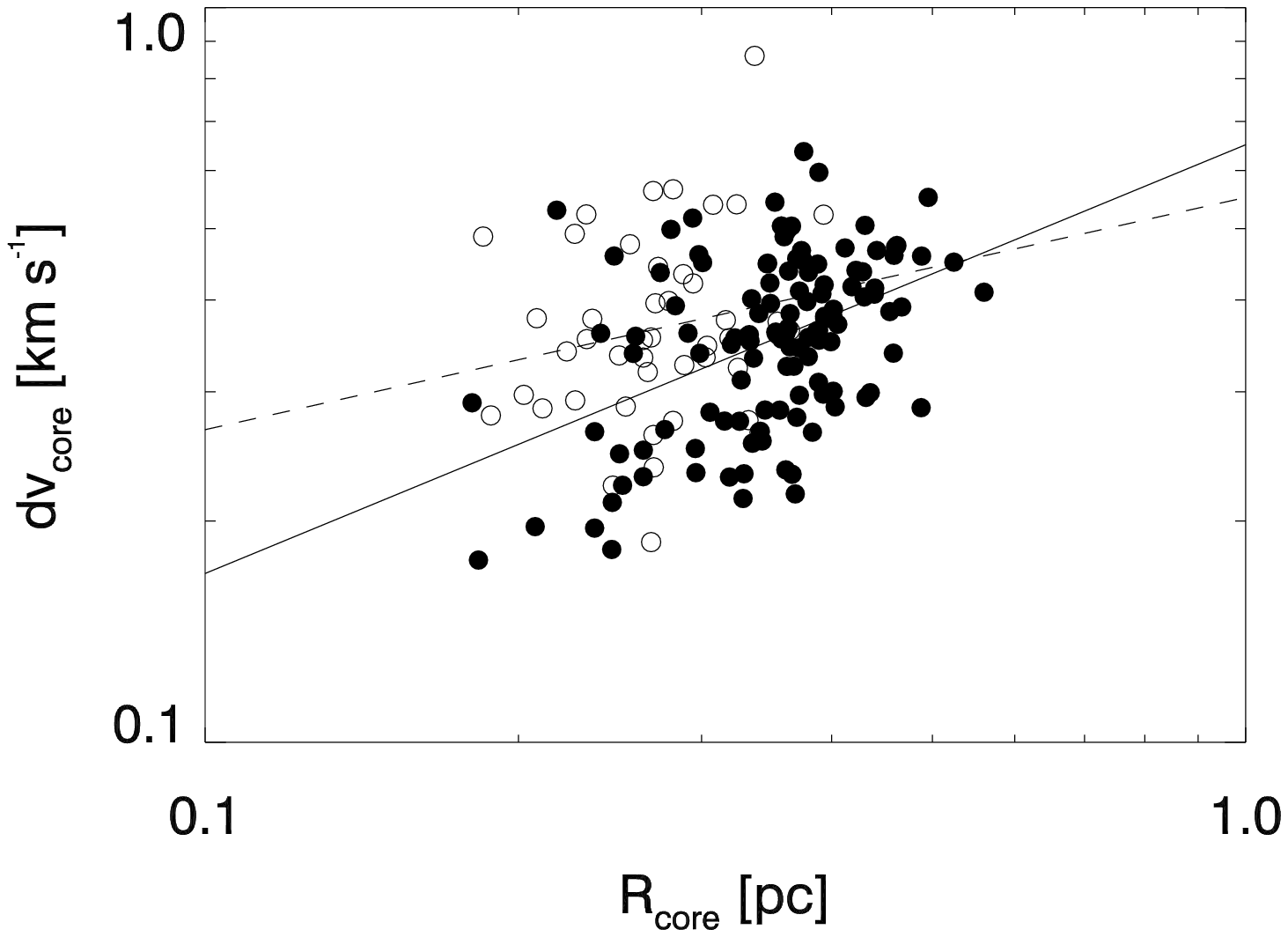}{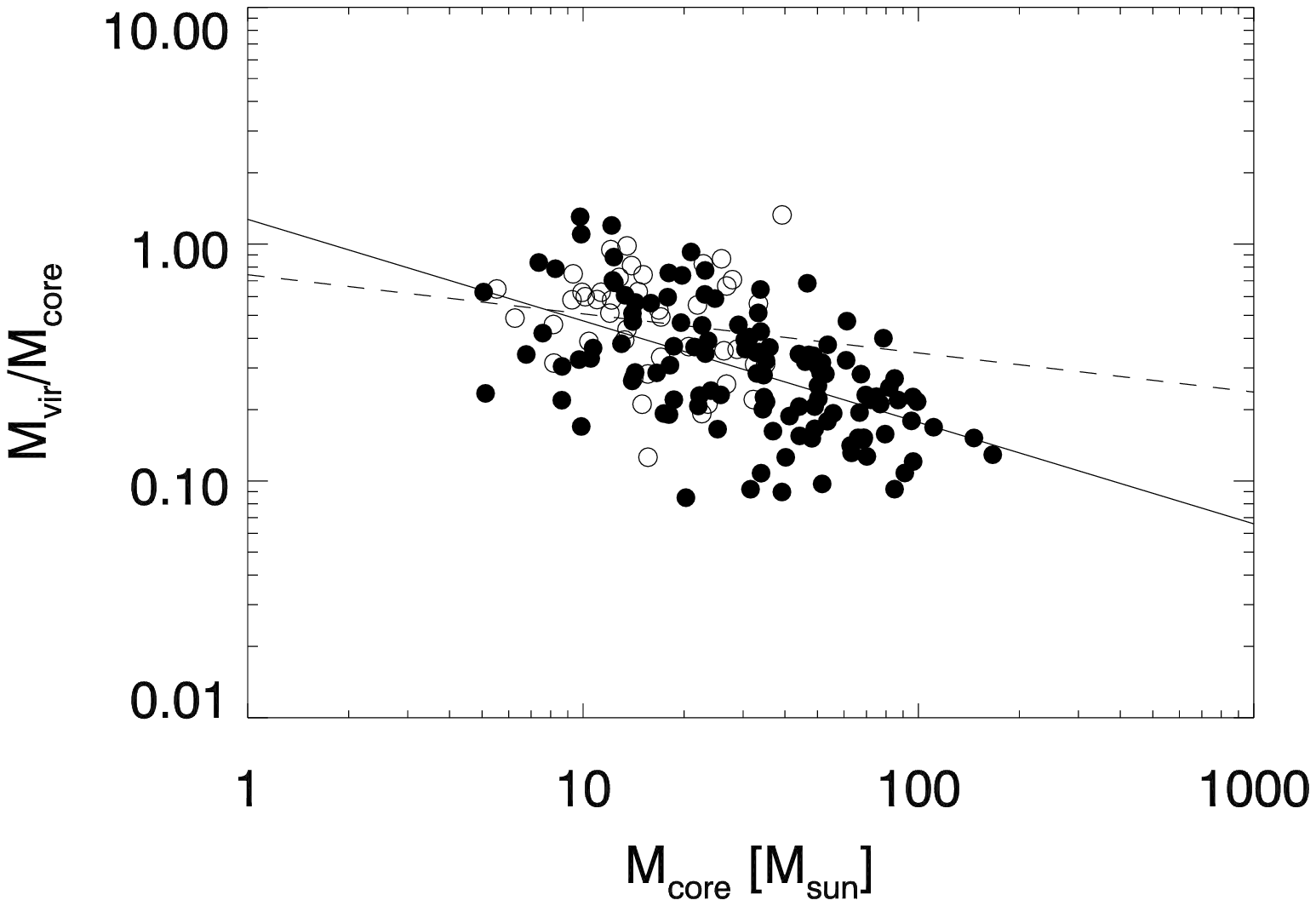}
\caption{
Velocity width-radius ({\it Left}) and virial ratio-mass ({\it Right}) relations of the {\CeighteenO} cores.
The filled and open circles correspond to the S 140 and the smoothed OMC-1 data, respectively.
The solid and dahsed lines show the best-fit power-law functions for the S140 and OMC-1 data, respectively
(see \S \ref{coreProperties}).
}
\label{corrPlots}
\end{figure}

In the left panel of Figure \ref{corrPlots}
we show the {\dvcore}-{\Rcore} relation of the {\CeighteenO} cores.
Although the distribution of the S140 cores tends to shift toward larger radii,
as shown in Figure \ref{histograms},
both the distributions in S140 and OMC-1 seem to be similar to each other.
Actually,
the best-fit power-law function for S140 is
({\dvcore}/{\kms}) = (0.65$\pm$0.07)$\times$({\Rcore}/pc)$^{0.58\pm0.11}$
with a small correlation coefficient of 0.39,
and the best-fit one for the OMC-1 cores is
({\dvcore}/{\kms}) = (0.55$\pm$0.21)$\times$({\Rcore}/pc)$^{0.32\pm0.25}$
with a correlation coefficient of 0.22;
the two best-fit functions are consistent with each other within the uncertainties.
In the right panel of Figure \ref{corrPlots}
we show the correlation between the virial ratio, {\Mvir}$/${\Mcore}, and {\Mcore}.
Since 
the mean value of the virial ratio is 0.3$\pm$0.2 in S140,
all the {\CeighteenO} cores are likely to be gravitationally bound
by considering the uncertainty in $X_{\rm C^{18}O}$ of a factor 3.
Both the data in S140 and OMC-1 show negative correlations;
the best-fit power-law function of the S140 cores is
({\Mvir}/{\Mcore}) = (1.27$\pm$0.25)$\times$({\Mcore}/{\Msun})$^{-0.43\pm0.06}$,
with a correlation coefficient of 0.48.
A similar negative correlation in the {\CeighteenO} \J{1}{0} data is also found
in low-mass star forming regions of Taurus, Ophiuchus, Lupus, L1333, Chamaeleon and the Pipe Nebula \citep{tac02}.
The power-law index of $-$0.43 is shallower than
the value of $-2/3$ expected for ``pressure-confined'' structures derived by \citet{ber92}.
This fact also suggests that
the self-gravity is dominant in the S140 {\CeighteenO} cores,
compared to the ambient pressure,
unlike the $\rho$ Ophiuchi case by \citet{mar10}.

\section{{\CeighteenO} CMF in the S140 region}
\label{mf_text}
\begin{figure}
\epsscale{1.0}
\plotone{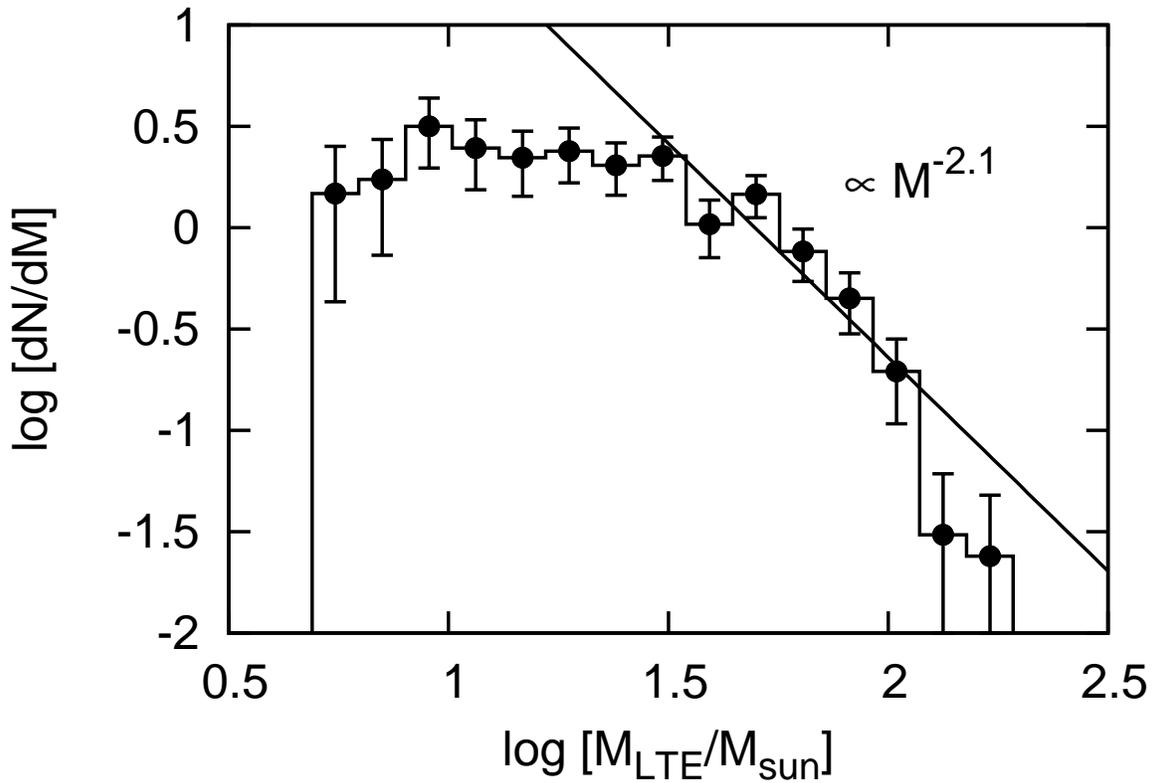}
\caption{
C$^{18}$O \J{1}{0} core mass function in the S 140 region.
The error bars show the statistical uncertainty of $\sqrt{N} $, 
where $N $ is the sample number in each mass bin.
The solid line indicates the best-fit power-law function
in the high-mass part above 30 {\Msun}.
}
\label{mf}
\end{figure}

\begin{figure}
\epsscale{1.0}
\plotone{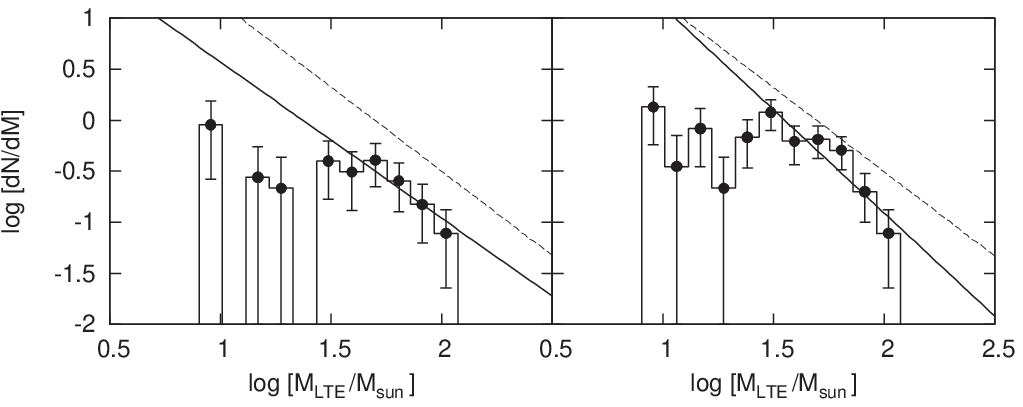}
\caption{
({\it Left}) Mass function of the C$^{18}$O cores
identified in the mapping region
by \citet{joh92} in the {\CeighteenO} \J{2}{1} line.
Note that the mass bins are the same as those in Figure \ref{mf}.
The solid line indicates the best-fit power-law function above 30 {\Msun},
and the dashed line a power-law function with the index of $-1.65$, 
which was derived from the {\CeighteenO} \J{2}{1} data by \citet{kra98}.
({\it Right}) Mass function of the {\CeighteenO} cores identified all over the head clump.
The solid line shows the best-fit power-law function above 30 {\Msun},
and the dashed one is the same as in the left panel. 
}
\label{mf_4x4}
\end{figure}

Figure \ref{mf} shows a CMF of the {\CeighteenO} cores. 
The CMF has a turnover at around 30 {\Msun},
and a power-law-like shape in the high-mass part above the turnover.
Above 30 $M_{\odot}$, we applied a single power-law function
by considering the statistical uncertainties
and found that the best-fit power-law index $\gamma$ is 2.1$\pm$0.2.
However,
the $\gamma$ value significantly differs from
that of the {\CeighteenO} \J{2}{1} CMF in the S140 region of $\gamma$ = 1.65$\pm$0.18 \citep{kra98}.
We consider that this difference is likely due to their limited observation area
in core survey.
As shown in Figure \ref{totalMap},
it is apparent that
the area of the {\CeighteenO} \J{2}{1} observations \citep{joh92} covers
only the central intense part of the head clump, where the {\CeighteenO} intensities are relatively
stronger than those in the other areas in the S140 region,
and the core identification is likely to be biased to massive cores.
To examine this possibility, 
we re-identifed 26 {\CeighteenO} \J{1}{0} cores within the {\CeighteenO} \J{2}{1} mapping area,
and we re-estimated the CMF from the 26 cores,
as shown in the left panel of Figure \ref{mf_4x4}.
Although our spatial resolution of 22$''$ is twice coarser than
that of the {\CeighteenO} \J{2}{1} study of 13$''$ \citep{joh92,kra98}
and hence our statistical uncertainties are large
in the low-mass part of $\leq$ 30 {\Msun},
the CMF shows a power-law-like behavior in the high-mass part.
The best-fit $\gamma$ is 1.5$\pm$0.6 and 
is consistent with the previous value of 1.65 derived by \citet{kra98}.
Furthermore, we estimated the CMF all over the head clump.
There are 44 cores in the head clump and the best-fit $\gamma$ is 2.0$\pm$0.4, 
as shown in the right panel of Figure \ref{mf_4x4}.
Although the core numbers are not sufficiently large for the statistical analysis,
the $\gamma$ value in the head clump tends to be larger than that in the central intense part of the head clump,
and are consistent with that of all the {\CeighteenO} cores in our observed region.
Therefore,
we conclude that
the CMF should be estimated from the mapping data
that covers at least one clump, which is thought to be the natal structure
of a stellar cluster \citep{lad03}.

\begin{figure}
\epsscale{1.0}
\plotone{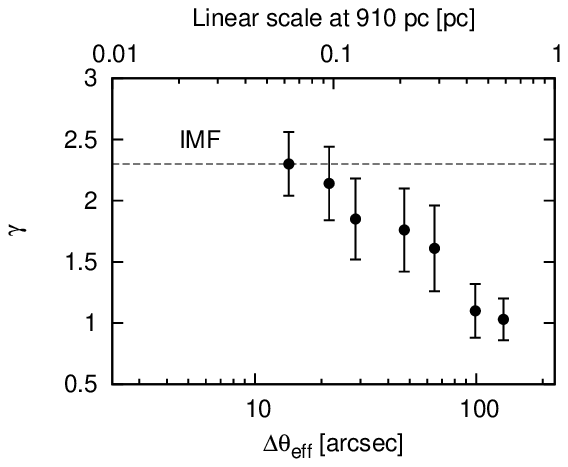}
\caption{
Power-law index $\gamma$ of the C$^{18}$O \J{1}{0} CMF,
derived from the smoothed data cube in the OMC-1 region \citep{ike09b},
is shown as a function of the effective spatial resolution,
\Resolution{eff}, in arcsec.
The upper x-axis is in pc at the distance to the S140 region.
The error bars indicate the uncertainties in fitting a power-law function
to the CMF in the high-mass part.
The horizontal dashed line shows the $\gamma$ value of the Salpeter IMF \citep{sal55}.
}
\label{gamma-res}
\end{figure}

Our $\gamma$ value of 2.1$\pm$0.2 is consistent with 
that of the {\CeighteenO} CMF in the OMC-1 region \citep{ike09b}.
This agreement confirms that our observations could resolve star-forming cores
even at the large distance of $\sim $ 1 kpc.
A poor spatial resolution is
one of the major causes of the underestimation of $\gamma$.
To examine the dependence of $\gamma$ on the spatial resolution,
we created the smoothed OMC-1 {\CeighteenO} data cubes
by changing the effective resolutions, as described in \S \ref{coreProperties},
and derived the $\gamma$ values for the smoothed cubes,
as shown in Figure \ref{gamma-res}.
It is clearly shown that
our resolution of 22$''$, corresponding to 0.097 pc for S140,
can correctly estimate the $\gamma $ value within the uncertainties.
Actually, our $\gamma$ value is consistent with
that in the {\CeighteenO} study by \citet{tac02},
having a spatial resolution enough to resolve 0.1 pc-scale cores. 
In addition, Figure \ref{gamma-res} predicts that
$\gamma$ is considerably underestimated for the case that
\Resolution{eff} becomes larger than 0.1 pc, which is the minimum radius of the cores in the OMC-1 region
at the highest resolution of 0.061 pc.
Therefore, the small $\gamma$ value of $\sim$ 1.7 in RCW 106 \citep{won08}
is likely due to the coarse spatial resolution of 0.78 pc at the distance of 3.6 kpc \citep{loc79}.

This study concludes that
the power-law shape with $\gamma$ $>$ 2 in CMF holds
even in tenuous structures with the densities of 10$^{3\mbox{\scriptsize --}4}$ {\pcc}
of the S140 region,
in addition to our recent work by \citet{ike09b} in OMC-1.
Furthermore,
the $\gamma$ values of the {\CeighteenO} CMFs in S140 and OMC-1 are quite consistent with 
that of the Galactic field-averaged IMF of 2.3$\pm$0.7 \citep{kro01a}.
These observational facts lead us to
the hypothesis that
the power-law nature in the IMF originates in molecular cloud structures
with densities of less than 
\sime{3\mbox{\scriptsize --}4} {\pcc}.

Our conclusion of the resemblance between the CMF and the IMF is
consistent with the recent theoretical works showing that
such a resemblance should be
understood as a statistical relation,
rather than a one-to-one correspondence between a core and a star to be formed within it.
\citet{smi09} examined the formation and evolution of gravitationally-bounded cores
by SPH simulations and found that the initial masses of the cores have
a poor correlation with the resultant stellar masses.
However, the shape of the CMF is maintained throughout the simulations 
and the resultant IMF shape is quite consistent with the initial shape of the CMF \citep[see also][]{cha10}.
Furthermore,
\citet{swi08} presented a supporting result that
the resultant IMFs are insensitive to the differences in theoretical models for star formation within cores
and that the input shape of the CMF is kept until the IMF
within the current observational uncertainties.

Since our studies are limited to the two clouds of S140 and OMC-1 up to now,
it is urgent to perform systematic studies of the tenuous CMFs in various star forming regions.
Since the S140 and OMC-1 regions are massive star/cluster forming ones,
it is interesting to explore the CMF shape in regions where no star formation activity is found
because stellar feedback processes, such as outflow, stellar wind, and expanding H {\sc ii} region
may cause some influence on the physical properties of the tenuous structures \citep[e.g.,][]{nak07}.
Although the global S140 cloud and the individual {\CeighteenO} cores
are likely to be influenced by the Cepheus Bubble as shown in \S \ref{map} and \ref{coreProperties},
we cannot examine whether the CMF shape is affected by the bubble or not 
only from our S140 data.
Recently,
\citet{rat09} estimated the DCMF in the Pipe Nebula,
where no active star formation occurs,
and found that the DCMF shape including the slope in the high-mass part is quite consistent with
that of the IMF
derived in the Orion Nebula Cluster associated with the OMC-1 cloud \citep{mue02},
where the most active star formations occur in the solar neighborhood.
This study suggests that
the stellar feedback processes seem not to play a dominant role in determining the CMF shape.

For future observations of the tenuous CMFs,
we require the following two points on the basis of our results.
First, mapping observations should not be spatially biased.
We have shown that the lower value of $\gamma $ in the {\CeighteenO} \J{2}{1} CMF
previously estimated in the S140 region is likely to be caused by
the limited mapping area only for the intense portion of the cloud.
Second, observations should be done with a high spatial resolution
enough to resolve the 0.1 pc-scale core structures, as shown in Figure \ref{gamma-res}. 
To further investigate the CMFs over the Galactic disk,
one should extend the studies in the solar neighborhood
to those in regions where environmental conditions, such as
metallicity, interstellar radiation, and turbulence, are different \citep{kru10,sch10}.
The first step would be toward the neighbor, Sagittarius/Perseus arms.
Actually, 
in M17SW and RCW 106 which are located in the Sagittarius arm,
smaller power-law indices of the CMFs of $\sim 1.7$ have been reported \citep{kra98,won08}.
To do observations of cores in the Sagittarius arms, approximately 2.0 kpc apart from the sun,
we need an angular resolution of 10$''$ to achieve a linear spatial resolution of 0.1 pc.
Furthermore, an extreme environment is known in the Galactic center region.
The Arches cluster, associated with the Galactic center, shows the present-day mass function
with $\gamma$ of $<2$, which is significantly flatter than that in the solar neighborhood \citep{sto05}.
To resolve the 0.1 pc at the distance to the Galactic center of 8 kpc,
we need a very high spatial resolution of 2$''$. 
ALMA is one of the best facilities to allow us to carry out
spatially unbiased observations with the spatial resolution as high as 0.1 pc
in the distant Galactic regions and
is expected to reveal the physical relationship between the CMF and the IMF in the Galaxy.


\section{Summary}
\label{summary}
We have carried out the mapping observations toward the S140 region
in the {\CeighteenO} \J{1}{0} line.
From the {\CeighteenO} map, which covers a $20'\times18'$ area in the region,
we construct the {\CeighteenO} core catalog.
Our results and conclusions are summarized as follows.

\begin{enumerate}
\item
We identified three clumps in the {\CeighteenO} map.
The head, tail, and filamentary clumps are likely to correspond to
the natal objects of the stellar clusters
because the clumps are in gravitationally bound state and
the masses of them is comparable to the typical one of cluster-forming clumps.
We found red-shifted components at the periphery of the head and tail clumps,
which can be interpreted as the pushed gas by the collision of the Cepheus Bubble. 

\item We identified 123 cores from the {\CeighteenO} cube data.
The radius and mass of the {\CeighteenO} cores in S140 tend to be larger than
those of the {\CeighteenO} cores in OMC-1 \citep{ike09b}.
The differences suggest that
the cometary cloud formed by the compression due to the collision of the expanding Cepheus Bubble.
On the other hand, the velocity widths of the {\CeighteenO} cores in the S140 and OMC-1 regions are
fairly consistent with each other,
possibly indicating that the turbulence excited by the bubble has already been dissipated.

\item We have demonstrated that
the power-law index in the high-mass part of CMF, $\gamma$,
could be underestimated due to insufficient spatial resolution.
Therefore, we should estimate the CMF with spatial resolutions
smaller than the minimum sizes of cores.
Furthermore, we have found that
the $\gamma$ value could also be biased if a mapping region is limited to a part of a clump.

\item The $\gamma$ values of the {\CeighteenO} CMFs in the S140 and OMC-1 regions
are quite consistent with 
that of the Galactic field-averaged IMF of 2.3$\pm$0.7 \citep{kro01a}.
Therefore, 
it is likely that
the power-law nature in the IMF originates in molecular cloud structures
with densities of less than 
\sime{3\mbox{\scriptsize --}4} {\pcc}.

\end{enumerate}

\acknowledgments

We thank A. Higuchi for providing useful information and data
to carry out our observations and to prepare the manuscript.
We are grateful to A. Yoshida and T. Akashi for providing {\tt VELSAN},
a nice software for calculating the centroid velocity/velocity width maps.
We also acknowledge an anonymous referee for valuable suggestions and comments that improve the paper.
For helping us in our observations and data reduction,
we are grateful to the staff of the Nobeyama Radio Observatory (NRO),
which is a branch of the National Astronomical Observatory of Japan,
National Institute of Natural Sciences.
This work is supported by
a Grant-in-Aid for Scientific Research (A) from
the Ministry of Education, Culture, Sports, Science and Technology of Japan
(No. 19204020).

\begin{deluxetable}{l l l l l l l c c c c c c c c c}
\tabletypesize{\scriptsize}
\rotate
\tablecaption{
Physical properties of the {\CeighteenO} cores in the S140 cloud
\label{propertyTable}
}
\tablewidth{0pt}
\tablehead{
\colhead{I.D.} &
\multicolumn{3}{c}{R.A.} &
\multicolumn{3}{c}{Decl.} &
\colhead{$v_{{\rm LSR}}$} &
\colhead{$T_{A,{\rm peak}}^*$} &
\multicolumn{2}{c}{$R_{\rm core}$} &
\colhead{$dv_{\rm core}$} &
\colhead{$M_{{\rm core}}$} &
\colhead{$M_{{\rm vir}}$} &
\colhead{$M_{{\rm vir}}$/$M_{{\rm core}}$} &
\colhead{$\bar{n}$}\\
\colhead{} &
\multicolumn{3}{c}{J2000.0} & 
\multicolumn{3}{c}{J2000.0} & 
\colhead{km s$^{-1}$} &
\colhead{K} &
\colhead{arcsec} & \colhead{pc} &
\colhead{km s$^{-1}$} &
\colhead{$M_{\odot}$} &
\colhead{$M_{\odot}$} &
\colhead{} &
\colhead{10$^{3}$ cm$^{-3}$}
}
\startdata
  1 & 22 & 19 & 16 & 63 & 18 & 57.0 & $-$7.6 & 1.71 &  82.1 &  0.36 &  0.50 & 111.0 &  18.7 & 0.2 & 9.8\\
  2 & 22 & 19 & 16 & 63 & 18 & 57.0 & $-$6.4 & 1.81 &  81.1 &  0.36 &  0.50 &  67.4 &  19.0 & 0.3 & 6.2\\
  3 & 22 & 19 & 16 & 63 & 18 & 57.0 & $-$5.1 & 0.68 &  49.3 &  0.22 &  0.53 &   9.8 &  12.8 & 1.3 & 4.0\\
  4 & 22 & 19 & 16 & 63 & 19 & 19.0 & $-$6.9 & 1.92 &  77.1 &  0.34 &  0.38 &  68.8 &  10.5 & 0.2 & 7.3\\
  5 & 22 & 19 & 19 & 63 & 20 & 47.0 & $-$8.4 & 0.73 &  81.1 &  0.36 &  0.35 &  33.0 &   9.4 & 0.3 & 3.0\\
  6 & 22 & 19 & 19 & 63 & 21 &  9.0 & $-$8.0 & 0.98 &  89.0 &  0.39 &  0.30 &  48.1 &   7.3 & 0.2 & 3.3\\
  7 & 22 & 19 & 22 & 63 & 18 & 13.0 & $-$6.8 & 2.01 &  79.1 &  0.35 &  0.42 &  66.8 &  13.0 & 0.2 & 6.6\\
  8 & 22 & 19 & 22 & 63 & 20 &  3.0 & $-$5.3 & 0.49 &  54.4 &  0.24 &  0.36 &   8.3 &   6.5 & 0.8 & 2.5\\
  9 & 22 & 19 & 22 & 63 & 20 & 25.0 & $-$7.3 & 1.24 &  69.2 &  0.31 &  0.28 &  40.2 &   5.1 & 0.1 & 5.9\\
 10 & 22 & 19 & 25 & 63 & 20 &  3.0 & $-$7.6 & 1.19 &  59.7 &  0.26 &  0.23 &  31.6 &   2.9 & 0.1 & 7.2\\
 11 & 22 & 19 & 25 & 63 & 20 & 25.0 & $-$6.7 & 0.79 &  80.1 &  0.35 &  0.36 &  34.5 &   9.7 & 0.3 & 3.3\\
 12 & 22 & 19 & 28 & 63 & 29 & 57.0 & $-$8.1 & 0.71 &  73.2 &  0.32 &  0.36 &  12.4 &   8.5 & 0.7 & 1.5\\
 13 & 22 & 19 & 29 & 63 & 19 & 41.0 & $-$8.5 & 1.20 &  97.1 &  0.43 &  0.44 &  95.3 &  17.1 & 0.2 & 5.1\\
 14 & 22 & 19 & 29 & 63 & 21 & 53.0 & $-$7.1 & 0.95 &  75.9 &  0.34 &  0.40 &  35.1 &  11.3 & 0.3 & 3.9\\
 15 & 22 & 19 & 29 & 63 & 24 &  5.0 & $-$8.3 & 0.75 &  83.9 &  0.37 &  0.46 &  47.1 &  16.0 & 0.3 & 3.9\\
 16 & 22 & 19 & 32 & 63 & 18 & 35.0 & $-$6.2 & 0.74 & 105.8 &  0.47 &  0.39 &  52.8 &  14.9 & 0.3 & 2.2\\
 17 & 22 & 19 & 32 & 63 & 19 & 19.0 & $-$8.0 & 1.28 &  97.8 &  0.43 &  0.29 &  84.9 &   7.8 & 0.1 & 4.4\\
 18 & 22 & 19 & 32 & 63 & 20 &  3.0 & $-$7.4 & 1.21 &  55.7 &  0.25 &  0.18 &  20.3 &   1.7 & 0.1 & 5.7\\
 19 & 22 & 19 & 32 & 63 & 20 & 25.0 & $-$7.1 & 1.53 &  77.4 &  0.34 &  0.26 &  51.6 &   5.0 & 0.1 & 5.4\\
 20 & 22 & 19 & 32 & 63 & 21 &  9.0 & $-$7.6 & 1.64 &  85.8 &  0.38 &  0.40 &  79.6 &  12.5 & 0.2 & 6.2\\
 21 & 22 & 19 & 32 & 63 & 21 & 31.0 & $-$8.4 & 1.19 &  78.7 &  0.35 &  0.45 &  51.1 &  14.6 & 0.3 & 5.1\\
 22 & 22 & 19 & 35 & 63 & 18 & 35.0 & $-$6.8 & 1.01 &  82.7 &  0.36 &  0.35 &  44.1 &   9.1 & 0.2 & 3.8\\
 23 & 22 & 19 & 35 & 63 & 18 & 35.0 & $-$6.4 & 0.97 &  83.6 &  0.37 &  0.22 &  33.9 &   3.7 & 0.1 & 2.8\\
 24 & 22 & 19 & 35 & 63 & 20 & 47.0 & $-$7.4 & 1.47 &  72.3 &  0.32 &  0.23 &  39.2 &   3.5 & 0.1 & 5.1\\
 25 & 22 & 19 & 35 & 63 & 29 & 35.0 & $-$7.8 & 0.65 &  56.7 &  0.25 &  0.25 &   7.6 &   3.2 & 0.4 & 2.0\\
 26 & 22 & 19 & 38 & 63 & 19 & 19.0 & $-$9.0 & 0.77 & 127.0 &  0.56 &  0.41 &  60.9 &  19.7 & 0.3 & 1.5\\
 27 & 22 & 19 & 38 & 63 & 19 & 19.0 & $-$7.5 & 1.27 &  86.1 &  0.38 &  0.33 &  70.2 &   8.9 & 0.1 & 5.4\\
 28 & 22 & 19 & 38 & 63 & 19 & 19.0 & $-$6.8 & 0.97 &  88.1 &  0.39 &  0.35 &  49.0 &  10.1 & 0.2 & 3.5\\
 29 & 22 & 19 & 38 & 63 & 25 & 11.0 & $-$8.6 & 1.23 &  85.2 &  0.38 &  0.64 &  46.6 &  31.9 & 0.7 & 3.7\\
 30 & 22 & 19 & 45 & 63 & 22 & 15.0 & $-$8.5 & 0.49 &  63.5 &  0.28 &  0.50 &  12.2 &  14.6 & 1.2 & 2.3\\
 31 & 22 & 19 & 45 & 63 & 23 & 21.0 & $-$7.0 & 0.75 &  83.9 &  0.37 &  0.28 &  25.7 &   5.9 & 0.2 & 2.1\\
 32 & 22 & 19 & 45 & 63 & 23 & 43.0 & $-$6.8 & 0.74 &  78.2 &  0.35 &  0.28 &  24.1 &   5.8 & 0.2 & 2.5\\
 33 & 22 & 19 & 45 & 63 & 28 & 29.0 & $-$7.9 & 1.12 &  66.0 &  0.29 &  0.36 &  21.5 &   7.9 & 0.4 & 3.7\\
 34 & 22 & 19 & 48 & 63 & 17 & 29.0 & $-$7.4 & 0.74 &  89.3 &  0.39 &  0.38 &  33.5 &  11.7 & 0.3 & 2.3\\
 35 & 22 & 19 & 48 & 63 & 20 &  3.0 & $-$7.9 & 1.26 &  86.1 &  0.38 &  0.36 &  66.1 &  10.0 & 0.2 & 5.1\\
 36 & 22 & 19 & 48 & 63 & 20 &  3.0 & $-$6.3 & 0.55 &  75.5 &  0.33 &  0.36 &  15.9 &   9.0 & 0.6 & 1.8\\
 37 & 22 & 19 & 48 & 63 & 21 & 53.0 & $-$6.8 & 0.72 &  91.0 &  0.40 &  0.39 &  50.2 &  12.7 & 0.3 & 3.3\\
 38 & 22 & 19 & 48 & 63 & 28 & 29.0 & $-$7.5 & 0.71 &  82.1 &  0.36 &  0.32 &  23.2 &   8.0 & 0.3 & 2.0\\
 39 & 22 & 19 & 50 & 63 & 20 & 25.0 & $-$8.5 & 1.99 &  93.4 &  0.41 &  0.47 &  87.0 &  19.1 & 0.2 & 5.2\\
 40 & 22 & 19 & 50 & 63 & 21 & 31.0 & $-$7.6 & 1.23 &  87.9 &  0.39 &  0.45 &  76.9 &  16.2 & 0.2 & 5.5\\
 41 & 22 & 19 & 50 & 63 & 22 & 59.0 & $-$7.4 & 1.12 & 110.5 &  0.49 &  0.29 &  63.2 &   8.3 & 0.1 & 2.3\\
 42 & 22 & 19 & 50 & 63 & 23 & 21.0 & $-$8.2 & 0.51 &  90.5 &  0.40 &  0.35 &  22.6 &  10.3 & 0.5 & 1.5\\
 43 & 22 & 19 & 50 & 63 & 24 &  5.0 & $-$6.3 & 0.71 &  76.3 &  0.34 &  0.33 &  34.6 &   7.8 & 0.2 & 3.8\\
 44 & 22 & 19 & 50 & 63 & 25 & 33.0 & $-$6.8 & 0.82 &  97.4 &  0.43 &  0.40 &  46.0 &  14.6 & 0.3 & 2.4\\
 45 & 22 & 19 & 50 & 63 & 27 & 45.0 & $-$7.9 & 0.97 &  58.7 &  0.26 &  0.36 &  18.6 &   6.9 & 0.4 & 4.5\\
 46 & 22 & 19 & 55 & 63 & 18 & 35.0 & $-$5.7 & 0.79 &  67.6 &  0.30 &  0.46 &  29.0 &  13.2 & 0.5 & 4.6\\
 47 & 22 & 19 & 55 & 63 & 22 & 59.0 & $-$7.0 & 1.00 &  82.7 &  0.36 &  0.38 &  50.3 &  11.2 & 0.2 & 4.4\\
 48 & 22 & 19 & 55 & 63 & 24 & 27.0 & $-$5.8 & 0.85 &  76.1 &  0.34 &  0.26 &  22.1 &   4.6 & 0.2 & 2.5\\
 49 & 22 & 19 & 55 & 63 & 25 & 33.0 & $-$6.4 & 1.05 &  88.0 &  0.39 &  0.31 &  41.3 &   7.7 & 0.2 & 3.0\\
 50 & 22 & 19 & 55 & 63 & 26 & 17.0 & $-$7.8 & 0.48 &  81.6 &  0.36 &  0.49 &  23.1 &  17.9 & 0.8 & 2.1\\
 51 & 22 & 19 & 58 & 63 & 19 & 19.0 & $-$6.4 & 0.98 &  88.1 &  0.39 &  0.60 &  61.2 &  28.9 & 0.5 & 4.4\\
 52 & 22 & 19 & 58 & 63 & 21 & 31.0 & $-$6.4 & 0.48 &  77.7 &  0.34 &  0.26 &  16.6 &   4.7 & 0.3 & 1.7\\
 53 & 22 & 19 & 58 & 63 & 23 & 43.0 & $-$7.8 & 0.76 &  90.9 &  0.40 &  0.30 &  35.1 &   7.6 & 0.2 & 2.3\\
 54 & 22 & 19 & 58 & 63 & 26 & 17.0 & $-$6.1 & 0.82 &  89.2 &  0.39 &  0.42 &  33.9 &  14.5 & 0.4 & 2.3\\
 55 & 22 & 19 & 58 & 63 & 27 & 45.0 & $-$8.4 & 0.71 &  83.0 &  0.37 &  0.23 &  18.7 &   4.1 & 0.2 & 1.6\\
 56 & 22 & 19 & 58 & 63 & 28 &  7.0 & $-$7.7 & 1.13 &  84.4 &  0.37 &  0.30 &  44.2 &   6.9 & 0.2 & 3.6\\
 57 & 22 & 19 & 58 & 63 & 29 & 35.0 & $-$8.6 & 0.76 &  53.7 &  0.24 &  0.26 &  10.5 &   3.5 & 0.3 & 3.3\\
 58 & 22 & 19 & 58 & 63 & 29 & 35.0 & $-$7.4 & 0.90 &  62.7 &  0.28 &  0.27 &  14.3 &   4.1 & 0.3 & 2.8\\
 59 & 22 & 20 &  1 & 63 & 19 & 41.0 & $-$8.3 & 1.12 &  84.4 &  0.37 &  0.41 &  35.9 &  13.2 & 0.4 & 2.9\\
 60 & 22 & 20 &  1 & 63 & 24 & 49.0 & $-$8.9 & 0.53 &  78.6 &  0.35 &  0.45 &  19.7 &  14.5 & 0.7 & 2.0\\
 61 & 22 & 20 &  1 & 63 & 25 & 11.0 & $-$8.0 & 1.10 &  84.4 &  0.37 &  0.46 &  51.5 &  16.3 & 0.3 & 4.2\\
 62 & 22 & 20 &  1 & 63 & 27 & 23.0 & $-$6.3 & 1.61 &  66.6 &  0.29 &  0.52 &  48.7 &  16.4 & 0.3 & 8.0\\
 63 & 22 & 20 &  1 & 63 & 28 & 29.0 & $-$8.8 & 0.72 &  74.2 &  0.33 &  0.31 &  14.1 &   6.6 & 0.5 & 1.7\\
 64 & 22 & 20 &  1 & 63 & 28 & 29.0 & $-$8.1 & 0.99 &  80.8 &  0.36 &  0.28 &  36.8 &   6.0 & 0.2 & 3.4\\
 65 & 22 & 20 &  5 & 63 & 19 & 19.0 & $-$5.3 & 0.59 &  62.0 &  0.27 &  0.44 &   9.9 &  10.9 & 1.1 & 2.0\\
 66 & 22 & 20 &  5 & 63 & 24 &  5.0 & $-$7.1 & 0.98 &  95.7 &  0.42 &  0.44 &  74.9 &  17.0 & 0.2 & 4.2\\
 67 & 22 & 20 &  5 & 63 & 26 & 39.0 & $-$8.0 & 1.38 &  82.3 &  0.36 &  0.37 &  68.4 &  10.2 & 0.1 & 6.0\\
 68 & 22 & 20 &  5 & 63 & 26 & 39.0 & $-$6.3 & 1.31 &  85.1 &  0.38 &  0.45 &  69.7 &  16.1 & 0.2 & 5.5\\
 69 & 22 & 20 &  5 & 63 & 28 & 51.0 & $-$6.5 & 0.73 &  91.0 &  0.40 &  0.38 &  31.0 &  11.9 & 0.4 & 2.0\\
 70 & 22 & 20 &  5 & 63 & 29 & 57.0 & $-$7.2 & 0.76 &  53.6 &  0.24 &  0.20 &   8.6 &   1.9 & 0.2 & 2.7\\
 71 & 22 & 20 &  8 & 63 & 20 & 47.0 & $-$7.2 & 0.49 & 104.7 &  0.46 &  0.47 &  33.8 &  21.7 & 0.6 & 1.4\\
 72 & 22 & 20 &  8 & 63 & 22 & 15.0 & $-$6.3 & 0.77 &  84.8 &  0.37 &  0.47 &  33.3 &  17.0 & 0.5 & 2.7\\
 73 & 22 & 20 &  8 & 63 & 29 & 57.0 & $-$6.8 & 0.91 &  73.9 &  0.33 &  0.27 &  22.2 &   5.1 & 0.2 & 2.7\\
 74 & 22 & 20 & 11 & 63 & 22 & 59.0 & $-$8.3 & 0.48 &  72.6 &  0.32 &  0.35 &  13.3 &   8.1 & 0.6 & 1.7\\
 75 & 22 & 20 & 11 & 63 & 23 & 21.0 & $-$7.7 & 1.02 &  88.1 &  0.39 &  0.36 &  55.7 &  10.7 & 0.2 & 4.0\\
 76 & 22 & 20 & 11 & 63 & 25 & 55.0 & $-$7.3 & 2.03 &  81.9 &  0.36 &  0.36 &  91.1 &   9.8 & 0.1 & 8.1\\
 77 & 22 & 20 & 11 & 63 & 26 & 39.0 & $-$7.1 & 2.35 & 110.6 &  0.49 &  0.46 & 166.6 &  21.5 & 0.1 & 6.0\\
 78 & 22 & 20 & 11 & 63 & 29 & 35.0 & $-$7.3 & 0.71 &  67.7 &  0.30 &  0.34 &  14.0 &   7.2 & 0.5 & 2.2\\
 79 & 22 & 20 & 14 & 63 & 26 & 39.0 & $-$8.5 & 1.02 &  94.8 &  0.42 &  0.42 &  51.2 &  15.2 & 0.3 & 2.9\\
 80 & 22 & 20 & 17 & 63 & 18 & 35.0 & $-$8.3 & 1.24 &  68.1 &  0.30 &  0.45 &  31.4 &  12.7 & 0.4 & 4.9\\
 81 & 22 & 20 & 17 & 63 & 23 & 21.0 & $-$6.5 & 0.48 &  84.4 &  0.37 &  0.34 &  23.6 &   9.2 & 0.4 & 1.9\\
 82 & 22 & 20 & 20 & 63 & 24 &  5.0 & $-$6.3 & 0.54 &  82.4 &  0.36 &  0.44 &  24.7 &  14.6 & 0.6 & 2.2\\
 83 & 22 & 20 & 20 & 63 & 25 & 33.0 & $-$7.3 & 2.18 &  79.9 &  0.35 &  0.54 &  96.3 &  21.8 & 0.2 & 9.2\\
 84 & 22 & 20 & 24 & 63 & 23 & 21.0 & $-$7.9 & 1.63 &  97.6 &  0.43 &  0.51 &  84.9 &  23.0 & 0.3 & 4.5\\
 85 & 22 & 20 & 28 & 63 & 24 &  5.0 & $-$6.5 & 0.51 &  57.1 &  0.25 &  0.22 &   8.7 &   2.6 & 0.3 & 2.3\\
 86 & 22 & 20 & 28 & 63 & 24 & 27.0 & $-$8.4 & 0.53 &  88.7 &  0.39 &  0.41 &  18.0 &  13.6 & 0.8 & 1.3\\
 87 & 22 & 20 & 28 & 63 & 28 &  7.0 & $-$6.0 & 0.52 &  83.4 &  0.37 &  0.32 &  14.3 &   8.1 & 0.6 & 1.2\\
 88 & 22 & 20 & 31 & 63 & 22 & 15.0 & $-$8.0 & 0.48 &  71.5 &  0.32 &  0.27 &  13.0 &   4.9 & 0.4 & 1.7\\
 89 & 22 & 20 & 31 & 63 & 26 & 17.0 & $-$7.7 & 0.86 &  81.9 &  0.36 &  0.23 &  25.2 &   4.2 & 0.2 & 2.2\\
 90 & 22 & 20 & 34 & 63 & 21 &  9.0 & $-$8.9 & 0.58 & 103.1 &  0.46 &  0.39 &  23.1 &  14.1 & 0.6 & 1.0\\
 91 & 22 & 20 & 34 & 63 & 24 &  5.0 & $-$6.7 & 0.56 &  74.5 &  0.33 &  0.21 &   9.7 &   3.2 & 0.3 & 1.2\\
 92 & 22 & 20 & 34 & 63 & 24 & 49.0 & $-$6.5 & 0.48 &  55.8 &  0.25 &  0.21 &   6.8 &   2.3 & 0.3 & 1.9\\
 93 & 22 & 20 & 34 & 63 & 25 & 33.0 & $-$7.8 & 0.49 &  67.0 &  0.30 &  0.25 &  10.7 &   3.9 & 0.4 & 1.7\\
 94 & 22 & 20 & 34 & 63 & 26 & 17.0 & $-$8.0 & 0.72 &  91.0 &  0.40 &  0.38 &  30.3 &  12.0 & 0.4 & 2.0\\
 95 & 22 & 20 & 34 & 63 & 26 & 39.0 & $-$7.4 & 0.75 &  79.2 &  0.35 &  0.40 &  32.9 &  11.4 & 0.3 & 3.2\\
 96 & 22 & 20 & 34 & 63 & 28 &  7.0 & $-$6.5 & 1.08 & 104.1 &  0.46 &  0.46 &  81.9 &  20.3 & 0.2 & 3.6\\
 97 & 22 & 20 & 36 & 63 & 17 & 51.0 & $-$8.2 & 0.59 &  87.7 &  0.39 &  0.36 &  17.9 &  10.7 & 0.6 & 1.3\\
 98 & 22 & 20 & 36 & 63 & 18 & 13.0 & $-$7.8 & 1.12 &  67.1 &  0.30 &  0.23 &  17.4 &   3.4 & 0.2 & 2.8\\
 99 & 22 & 20 & 36 & 63 & 19 & 41.0 & $-$7.5 & 1.24 &  86.1 &  0.38 &  0.44 &  44.0 &  15.1 & 0.3 & 3.4\\
100 & 22 & 20 & 36 & 63 & 24 &  5.0 & $-$7.1 & 0.59 &  86.9 &  0.38 &  0.26 &  18.2 &   5.6 & 0.3 & 1.4\\
101 & 22 & 20 & 36 & 63 & 24 & 49.0 & $-$6.8 & 0.49 &  74.7 &  0.33 &  0.23 &  14.0 &   3.7 & 0.3 & 1.6\\
102 & 22 & 20 & 36 & 63 & 26 & 17.0 & $-$6.3 & 0.75 &  89.2 &  0.39 &  0.38 &  34.5 &  11.9 & 0.3 & 2.4\\
103 & 22 & 20 & 40 & 63 & 19 & 41.0 & $-$9.3 & 1.27 & 100.1 &  0.44 &  0.47 &  53.6 &  20.1 & 0.4 & 2.6\\
104 & 22 & 20 & 41 & 63 & 22 & 15.0 & $-$8.5 & 0.49 &  58.4 &  0.26 &  0.34 &   7.4 &   6.2 & 0.8 & 1.8\\
105 & 22 & 20 & 41 & 63 & 22 & 15.0 & $-$7.7 & 0.72 & 112.2 &  0.50 &  0.55 &  78.6 &  31.5 & 0.4 & 2.7\\
106 & 22 & 20 & 41 & 63 & 25 & 55.0 & $-$6.2 & 0.78 &  82.9 &  0.37 &  0.50 &  21.0 &  19.4 & 0.9 & 1.8\\
107 & 22 & 20 & 41 & 63 & 27 & 45.0 & $-$7.5 & 0.74 &  64.1 &  0.28 &  0.39 &  19.6 &   9.1 & 0.5 & 3.6\\
108 & 22 & 20 & 44 & 63 & 18 & 35.0 & $-$7.5 & 1.00 &  56.0 &  0.25 &  0.46 &  12.4 &  10.9 & 0.9 & 3.4\\
109 & 22 & 20 & 44 & 63 & 19 & 41.0 & $-$9.2 & 1.20 &  91.3 &  0.40 &  0.29 &  34.4 &   6.9 & 0.2 & 2.2\\
110 & 22 & 20 & 44 & 63 & 23 & 21.0 & $-$7.0 & 0.54 & 103.9 &  0.46 &  0.34 &  30.5 &  11.0 & 0.4 & 1.3\\
111 & 22 & 20 & 44 & 63 & 24 & 49.0 & $-$7.5 & 0.96 &  99.6 &  0.44 &  0.41 &  50.3 &  15.2 & 0.3 & 2.5\\
112 & 22 & 20 & 47 & 63 & 29 & 57.0 & $-$7.6 & 0.65 &  40.9 &  0.18 &  0.29 &   5.1 &   3.2 & 0.6 & 3.6\\
113 & 22 & 20 & 50 & 63 & 19 & 19.0 & $-$8.9 & 1.10 &  98.7 &  0.44 &  0.30 &  49.1 &   8.1 & 0.2 & 2.5\\
114 & 22 & 20 & 54 & 63 & 17 & 51.0 & $-$8.6 & 1.97 & 118.8 &  0.52 &  0.45 & 146.4 &  22.2 & 0.2 & 4.3\\
115 & 22 & 20 & 57 & 63 & 17 &  7.0 & $-$9.4 & 2.11 &  99.7 &  0.44 &  0.42 &  72.0 &  15.8 & 0.2 & 3.6\\
116 & 22 & 20 & 57 & 63 & 27 &  1.0 & $-$7.4 & 0.50 &  75.7 &  0.33 &  0.35 &  12.3 &   8.6 & 0.7 & 1.4\\
117 & 22 & 21 &  6 & 63 & 16 & 45.0 & $-$8.9 & 1.67 &  91.9 &  0.41 &  0.37 &  96.4 &  11.6 & 0.1 & 6.1\\
118 & 22 & 21 &  6 & 63 & 18 & 35.0 & $-$9.2 & 1.75 & 104.4 &  0.46 &  0.47 &  99.1 &  21.4 & 0.2 & 4.3\\
119 & 22 & 21 & 16 & 63 & 14 & 55.0 & $-$8.9 & 1.18 &  59.8 &  0.26 &  0.25 &  18.0 &   3.4 & 0.2 & 4.1\\
120 & 22 & 21 & 16 & 63 & 16 &  1.0 & $-$9.0 & 1.72 &  75.2 &  0.33 &  0.36 &  62.8 &   8.9 & 0.1 & 7.2\\
121 & 22 & 21 & 20 & 63 & 15 & 39.0 & $-$8.7 & 1.67 &  80.2 &  0.35 &  0.36 &  53.5 &   9.6 & 0.2 & 5.1\\
122 & 22 & 21 & 36 & 63 & 14 & 11.0 & $-$9.3 & 1.19 &  41.5 &  0.18 &  0.18 &   5.1 &   1.2 & 0.2 & 3.5\\
123 & 22 & 21 & 39 & 63 & 14 & 11.0 & $-$8.9 & 1.01 &  47.0 &  0.21 &  0.20 &   9.9 &   1.7 & 0.2 & 4.6\\

\enddata
{\small
\tablecomments{
The typical uncertainty of each quantity is as follows:
For $R_{\rm core}$, 0.04 pc,
derived from the uncertainty in the estimation of the core projected area.
For $dv_{\rm core}$, 0.13 km s$^{-1}$,
corresponding to the velocity resolution.
For $M_{\rm core}$, a factor of 3 (see text).
For $M_{\rm vir}$, a factor of 3,
derived from the uncertainties in $R_{\rm core}$ and $dv_{\rm core}$.
For $M_{\rm vir}/M_{\rm core}$, a factor of 4,
derived from the uncertainties in $M_{\rm core}$ and $M_{\rm vir}$.
For $\bar{n}$, a factor of 4,
derived from the uncertainties in $R_{\rm core}$ and $M_{\rm core}$.
}}
\end{deluxetable}

\end{document}